# Updates on the background estimates for the X-IFU instrument onboard of the ATHENA mission

S. Lotti[a], C. Macculi[a], M. D'Andrea[a], L. Piro[a], S. Molendi[b], F. Gastaldello[b], T. Mineo[c], A. D'Ai[c], A. Bulgarelli[d], V. Fioretti[d], C. Jacquey[e], M. Laurenza[a], P. Laurent[f]

[a]Istituto di Astrofisica e Planetologia Spaziali, Via fosso del cavaliere 100, Roma, Italy;
[b]INAF -IASF Milano, Via E. Bassini 15, Milano, Italy;
[c]IASF Palermo, Via Ugo La Malfa 153, 90146 Palermo, Italy;
[d]IASF Bologna, Via Piero Gobetti 101, 40129 Bologna, Italy;
[e]IRAP-CDPP, 9 avenue du colonel Roche, 31068 Toulouse, France;
[f]CEA/DSM/IRFU/SAp, Bat. 709 Orme des Merisiers, CEA Saclay, 91191 France

**ABSTRACT**

ATHENA is the second large mission in ESA Cosmic Vision 2015-2025, with a launch foreseen in 2028 towards the L2 orbit. The mission addresses the science theme "The Hot and Energetic Universe", by coupling a high-performance X-ray Telescope with two complementary focal-plane instruments. One of these, the X-ray Integral Field Unit (X-IFU) is a TES based kilo-pixel array, providing spatially resolved high-resolution spectroscopy (2.5 eV at 6 keV) over a 5 arcmin FoV.

The background for this kind of detectors accounts for several components: the diffuse Cosmic X-ray Background, the low energy particles (< ~100 keV) focalized by the mirrors and reaching the detector from inside the field of view, and the high energy particles (> ~100 MeV) crossing the spacecraft and reaching the focal plane from every direction. In particular, these high energy particles lose energy in the materials they cross, creating secondaries along their path that can induce an additional background component.

Each one of these components is under study of a team dedicated to the background issues regarding the X-IFU, with the aim to reduce their impact on the instrumental performances. This task is particularly challenging, given the lack of data on the background of X-ray detectors in L2, the uncertainties on the particle environment to be expected in such orbit, and the reliability of the models used in the Monte Carlo background computations. As a consequence, the activities addressed by the group range from the reanalysis of the data of previous missions like XMM-Newton, to the characterization of the L2 environment by data analysis of the particle monitors onboard of satellites present in the Earth magnetotail, to the characterization of solar events and their occurrence, and to the validation of the physical models involved in the Monte Carlo simulations. All these activities will allow to develop a set of reliable simulations to predict, analyze and find effective solutions to reduce the particle background experienced by the X-IFU, ultimately satisfying the scientific requirement that enables the science of ATHENA.

While the activities are still ongoing, we present here some preliminary results already obtained by the group. The L2 environment characterization activities, and the analysis and validation of the physical processes involved in the Monte Carlo simulations are the core of an ESA activity named AREMBES (Athena Radiation Environment Models and Effects), for which the work presented here represents a starting point.

# 1. INTRODUCTION

ATHENA[1] is an observatory class mission, whose launch is foreseen in 2028, that will be placed in orbit around the second Lagrangian point of the Sun-Earth system in a large halo orbit. The mission will address two key questions: how does ordinary matter assemble into the large scale structures that we see today, and how do black holes grow and shape the Universe. The mission includes two focal plane detectors: a Wide Field Imager[2] (WFI), and the one we will deal with in this paper: the X-ray Integral Field Unit[3] (X-IFU).

X-IFU is an array of 3840 Transition Edge Sensors (TES) 249 μm pitch, composed of Mo/Au sensors and 1-2.5 μm Au and 3-6 μm Bi absorbers that operates at cryogenic temperatures to achieve the high spectral resolution of 2.5 eV at 6 keV. The absorbers will be arranged in a hexagonal shape, covering the 5' equivalent diameter Field of View (FoV).

The particle background experienced by an X-ray instrument in orbit is dominant over the diffuse component above 2-3 keV[4]. In stationary conditions this background is induced by two families of particles: the high energy cosmic rays (E>150 MeV), which have sufficient energy to cross the spacecraft and reach the detector from every direction, and the low energy particles (E< few 100s keV, mostly of solar origin), that are concentrated by the optics and reach the detector from inside its Field of View.

Since no X-ray mission has ever flown in the L2 environment we estimate the background induced by both components using Monte Carlo simulations. In this paper we describe the state of the art of the L2 environment characterization and the particle background simulations for the X-IFU instrument.

## 1.1 Background induced by high energy particles

High energy particles will cross the spacecraft and reach the detector from every direction, generating secondary particles along their way that can induce further background. This background component is reduced with the use of the CryoAC detector and of a Kapton shield for secondary electrons, and is strongly influenced by materials and masses in the detector proximity. We estimate its contribution with Geant4 simulations, reproducing the L2 environment and the ATHENA mass model (Figure 1).

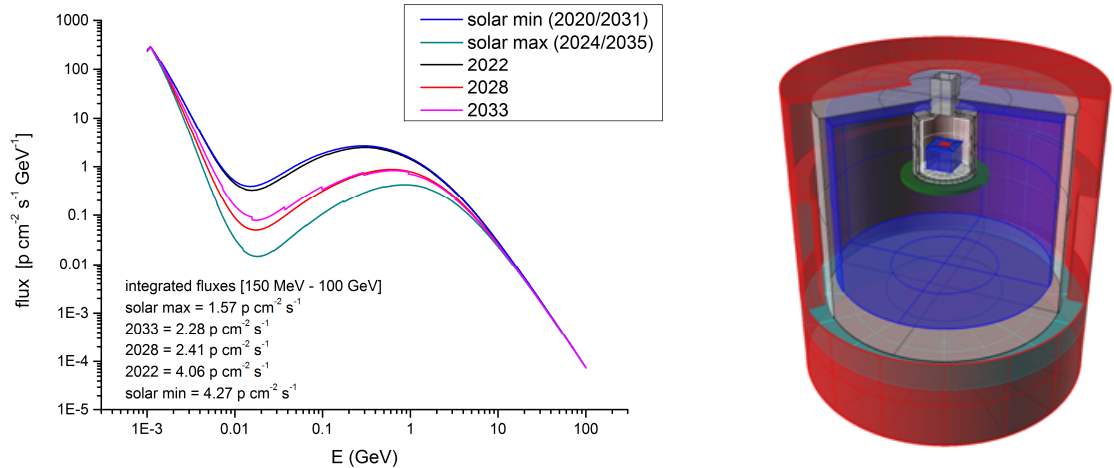

Figure 1. The fluxes of cosmic rays expected in L2 for different years in a solar cycle (left), and the preliminary mass model of the cryostat used in the simulations (right).

The first estimates were obtained using a simplified mass model for the Cryostat and the Focal Plane Assembly (FPA), due to the lack of information suffered in early stages of this work. However, with the mission progressing new information became available and we were able to upgrade the FPA mass model (Figure 2) and the Geant4 version, from 9.4 to 10.1. At the same time a complete revision of the Geant4 settings was performed. In the new configuration the background level obtained is $10^{-2}\ cts\ cm^{-2}\ s^{-1}$ in the 2-12 keV energy band (Figure 3).

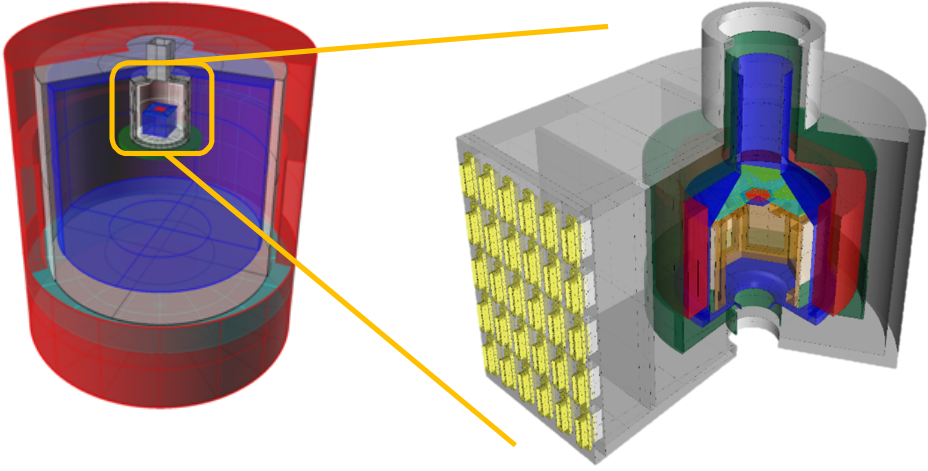

Figure 2. The preliminary cryostat mass model used to evaluate background; inside the yellow box the old model of FPA (left). The new FPA mass model used in the current simulations (right).

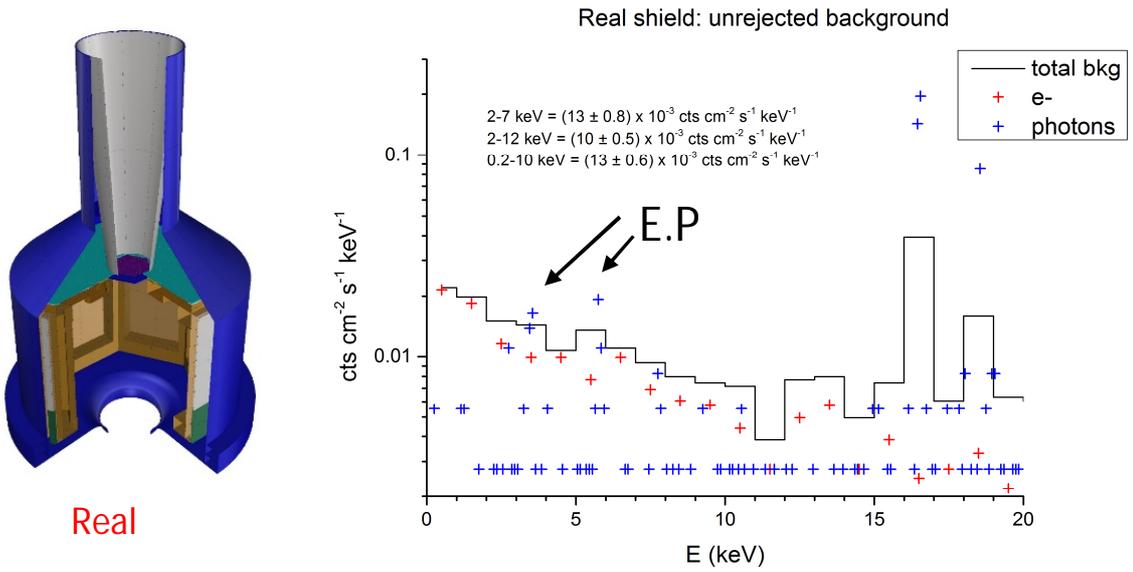

Figure 3. A detail of the new FPA: in blue the Niobium shield, in grey the 250 mm thick Kapton passive electron shield (left). On the right, the background expected in this configuration: the black line is the total (1 keV bin), red crosses represents the electrons contribution (1 keV bin), and the blue crosses the photons contribution (0.1 keV bin).

This background level was reached using the CryoAC detector, a secondary electron shield (Kapton, 250 μm thick) in the configuration shown in Figure 3, and pattern/energy selection criteria for the background events. As it can be seen from Figure 3 the main components of the background are secondary electrons (~75%, mostly coming from the surfaces directly seen by the detector), and secondary photons (~20%, half of which in the form of fluorescence lines from the niobium shield).

Starting from this result we addressed the two major components of the background: the secondary photons and electrons.

**Secondary photons:**

Roughly half of the secondary photons component comes from 16 and 18 keV fluorescence lines produced inside the Nb shield, when these photons impact the detector they induce the emission of 10.8 keV and 13 keV fluorescence photons from the Bi absorber. The 10.8 keV and 13 keV photons escape, leaving inside the detector a fixed amount of energy in the form of escape peaks. The remaining contribution is given by low energy photons that are completely absorbed and by high energy photons that Compton scatter in the detector, leaving a small fraction of their energy.

To block these 16 and 18 keV photons, we tested several configurations of double/tri-layered passive shieldings (see table 1).

Table 1. Background levels expected with different passive shielding configurations. The last line reports the presence of fluorescence lines or Escape Peaks in the spectrum

| Shield<br>bkg<br>2-12 keV | 250 um Kapton | 250 um Kapton + 10 um W | 250 um Kapton + 20 um Bi + 10 um W | 250 um Kapton + 250 um SiC + 10 um W | 250 um Kapton + 300 um $Si_3N_4$ + 10 um W | 10 um W + 300 um $Si_3N_4$ | 250 um Kapton + 1.3 mm $Si_3N_4$ | 250 um Kapton + 20 um Bi | 250 um Kapton + 1 um SiC |
|---|---|---|---|---|---|---|---|---|---|
| Total [x $10^{-3}$] cts/cm$^2$/s/keV | 10 | 7.6 | 8.8 | 8.4 | 8.1 | 7.8 | 7.4 | 8 | 7.3 |
| gamma [x $10^{-3}$] cts/cm$^2$/s/keV | 2 | 1.7 | 1.7 | 1.4 | 1.3 | 1 | 1.2 | 1.9 | 1.4 |
| Lines? | E.P. | W: 8.4 keV 9.6 keV | Bi: 10.8 keV | No | No | Si: 1.72 keV | No | Bi: 10.8 keV | E.P. 5.7 keV |

Summarizing the results obtained, we found that:

- A few μm of W placed between the Nb and the Kapton (a Kapton-W bilayer shield) efficently blocks Nb lines, but the Tungsten produces further fluorescences inside the instrument energy band.

- Inserting a further layer of Bi to block these W fluorescences (a Kapton-Bi-W tri-layer shield) we eliminate the W lines, but we get L fluorescences from Bi in turn;
    - Substituting Bi with SiC or $Si_3N_4$ we get rid of all fluorescences
    - Taking out Kapton (last surface: $Si_3N_4$) we have no fluorescences above 2 keV and a low background. The 1.72 keV line from Si can however be a hindrance for the observation of AGNs at redshift ~2-3

- Other bi-layers tested: Kapton-Bi, kapton-$Si_3N_4$, Kapton-SiC
    - The best result was obtained using a bilayer made of 250 μm of Kapton and 1.3 mm of $Si_3N_4$, roughly halving the photon component and reducing the total background by ~25%
    - The Kapton-Bi: bilayer also brought a ~20% background reduction. Furthermore, half of the photon background is concentrated in the Bi line at 10.8 keV, near the edge of the sensitivity band of the instrument. This solution is remarkable also since we already know it is feasible to cool down the Bi to cryogenic temperatures.
    - The Kapton-SiC solution brought results similar to the previous 2, but with an escape peak at 5.7 keV

**Secondary electrons**

Secondary electrons constitute the greatest contribution to the unrejected background. They have energies up to ~1 MeV and impact the detector surface with skew trajectories, backscattering and depositing only a fraction of their energy, not reaching the CryoAC detector.

In Figure 4 is shown the cumulative distribution of the spectrum of electrons that impact the detector depositing energies inside its sensitivity band. It can be seen that roughly 40% of the backscattered electrons impact the detector with E>100 keV and up to few MeV.

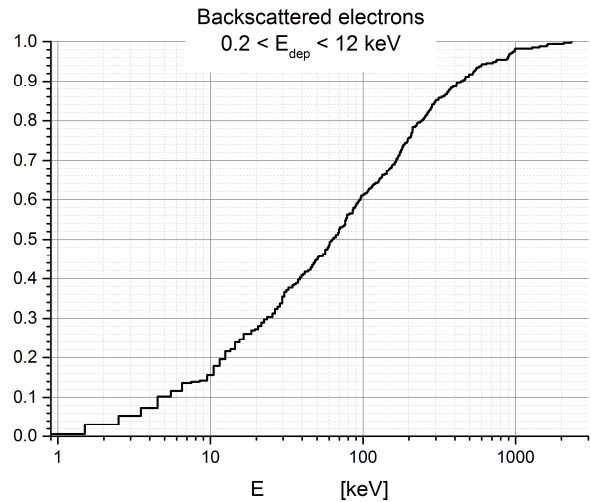

Figure 4. Cumulative distribution of the energies of secondary electrons that backscatter on the detector surface depositing energy inside its sensitivity band.

These electrons are difficult to block due to their high energies. We tested two approaches:

- substituting the lowest section of the Kapton shield with a higher density material like Tungsten, in the form of a 1 cm high ring with 1 mm thickness. With this composite shield we increase the stopping power for high energy electrons impacting the detector with small incidence angles. Tungsten however has a higher secondary electron yield, so it is not guaranteed that this approach would bring appreciable background reduction.
- inserting an electron filter just above the detector, to induce backscattering there and shield the detector from these energy releases. The filter thickness should be high enough to block electrons, providing at the same time acceptable X-ray transmission.

*Composite shield*

The background obtained with this configuration is shown in Figure 5, together with the background obtained with the "real" shield. The integrated background values for the main components are shown separately in the table. As it can be seen from the table this shield configuration allowed a background reduction of ~20%, mostly on the electron component that was reduced by ~30%. The photon component stayed constant, and exhibited an emission line at 8.4 keV from the Tungsten shield that accounted for $5 \times 10^{-4}$ cts/cm$^2$/s/keV in the 2-12 keV band.

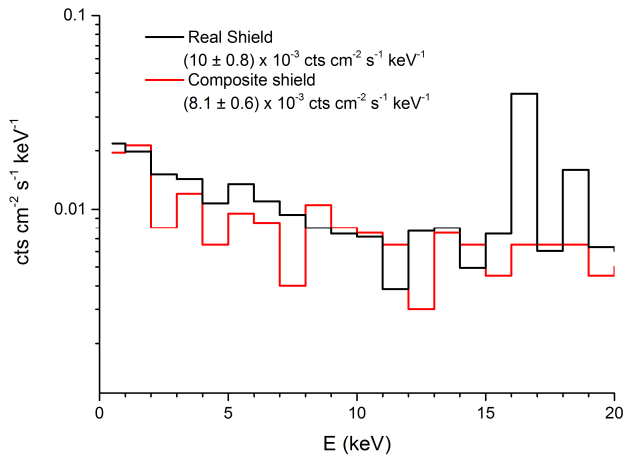

Figure 5. Background foreseen in the two configurations discussed in the text: in black the background with the 250 um Kapton shield, in red the one obtained with the composite shield. On the right a table that highlights the different contributions in the two configurations.

*Electron filter*

If we insert a filter just above the detector, a fraction of the secondary electrons will be backscattered there and not in the detector. Such filter needs to be thicker than the backscattering depth but thin enough to be transparent to X-rays.

From the extrapolation of experimental data[5] we estimate that 500 nm of Silicon should be able to block electrons up to ~80 keV. This thickness is quite high with respect to X-ray transparency, so we investigate the dependence of the backscattering depth with the atomic number[6] and derived some alternative configurations. In total 4 different configurations were tested:

- 500 nm Al: this allows to block electrons up to 82 keV
- 50 nm Au: same blocking power of 500 nm Al, higher Z
- 90 nm Al: lower stopping power, higher X-ray transmission
- 3 um BCB: this is to reproduce a configuration tested by the WFI team

In Figure 6 is reported the X-ray transmission of such filters. We found that there is no significant reduction of the backscattered electrons component using Al filters, while we obtain a ~20% background reduction above 2 keV using 50 nm Au, or 3 um BCB, however the X-ray transmission of such filters is too low to be considered for implementation.

The current level of confidence in these results is however low, pending a validation of the backscattering process (along with many others) to be performed inside the AREMBES framework.

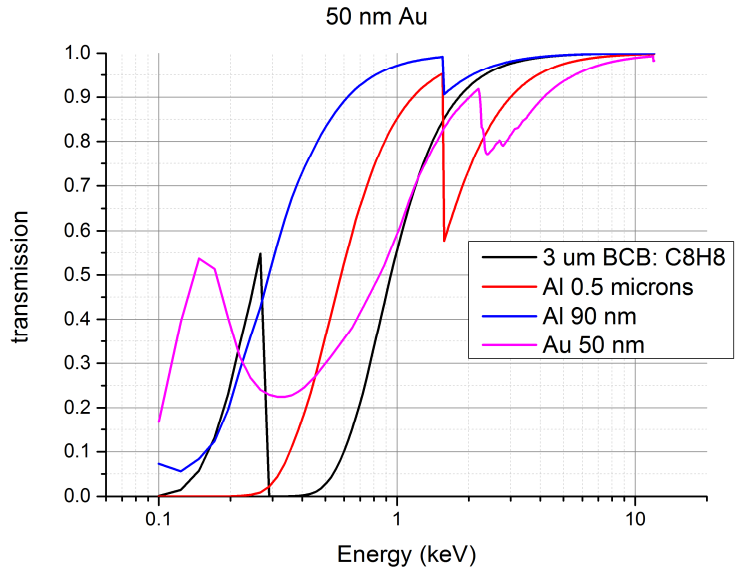

Figure 6. X-ray transmission of the different electron filters analyzed.

**1.2 Background induced by low energy particles**

Past X-ray missions like XMM-Newton and Chandra experienced sudden and intense increases of the background level (see Figure 7). Those "flares" were totally unrelated to the background rates measured by the radiation monitors, and were associated to protons of energies in the few tens to few hundreds of keV concentrated by the optics towards the focal plane. These low energy particles limited the exploitation of the data in two significant ways: firstly by reducing exposure times by a non-trivial fraction (about 40-50%, see Figure 7 - right) due to flaring events; secondly by contaminating the remaining data with a subdominant but poorly reproducible background component induced by the quiescent flux of these low energy particles.

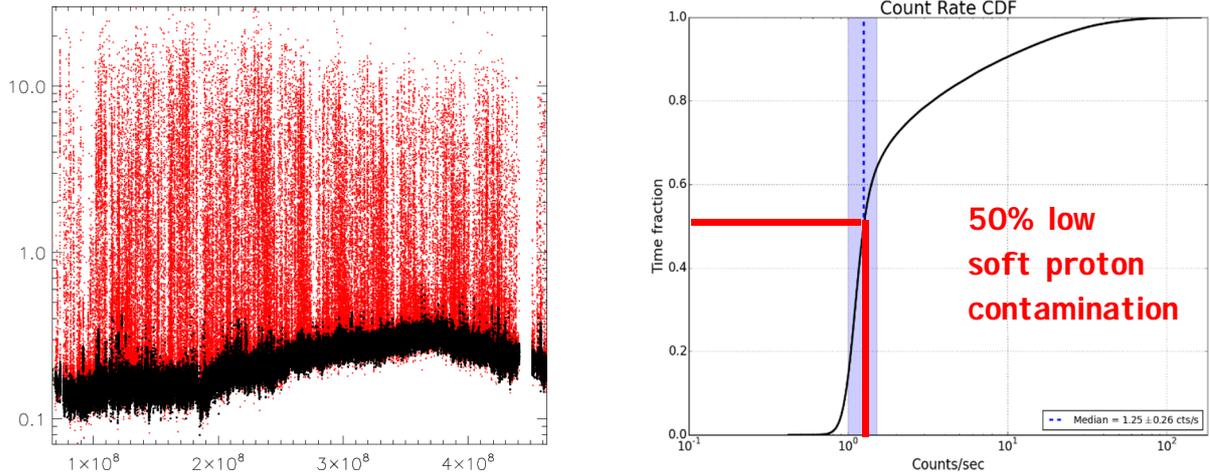

Figure 7. Count rates measured by EPIC/MOS during 100 Ms. Cosmic Rays are in black: they are slowly varying, exhibit modest spectral variations, and have high reproducibility (few %). Soft protons are in red, are highly variable in both spectrum and flux (left). On the right, the fraction of time the measured count rate is below a given value.

In a similar way low energy particles will enter the Athena SPO mirror aperture, scatter inside the mirrors and will be concentrated towards the focal plane. The effect will be even more significant given the larger collection area of the ATHENA mirrors. This soft proton flux must be blocked or diverted to avoid excess background loading on the WFI or X-IFU instruments.

The usual solution is to introduce a high magnetic field behind the rear mirror aperture, using magnets to defect the protons, preventing them from reaching the detectors. So far, all existing missions have used permanent magnets for this purpose, but feasibility studies to use superconductive magnets are under development. Typical magnetic fields involved are of the order of 100s G. These values are not sufficient to reflect backwards protons with E ~ keV, so it is likely that all the protons entering the mirrors will remain in the telescope tube.

One possible configuration for the diverter is the following. The permanent magnets are rectangular blocks with the magnetic field running across the thinnest dimension, that can be arranged either radially in the azimuthal gaps between the modules or azimuthally in the radial gaps between the rings of modules. The magnets alone (no support structure) are expected to absorb 8.3% of the incoming protons[7].

The scientific requirement for the ATHENA focal plane instruments to achieve a low energy particles induced background <10% of the GCR induced one. This translates into a residual flux at focal plane $< 0.1 \times B_{GCR} = 5 \times 10^{-3}\ p\ cm^{-2}\ s^{-1}$.

To calculate the expected residual flux at focal plane we can break down the particles interaction with the satellite in different modules (see Figure 8)

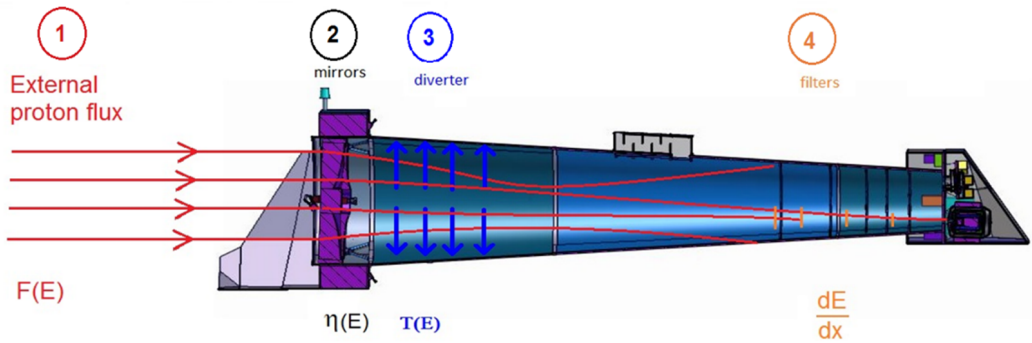

Figure 8. Schematics of the steps followed by the low energy particles: 1) the external proton flux F(E), 2) its interaction with the mirrors with focalization efficiency η, 3) deflection by the magnetic diverter with transmission efficiency T, 3) energy loss dE/dx inside the radiation filters

**External fluxes**

The low energy L2 environment is currently poorly known, complex, and highly dynamical (see Figure 9) we can assume as order of magnitude approximation the following fluxes at 80 keV for the different zones of the magnetotail assuming $F \propto E^{-1.5}$:

- Plasma sheet: $35 - 353\ p\ cm^{-2} s^{-1} sr^{-1} keV^{-1}$
- Lobes: $0.35 - 3.5\ p\ cm^{-2} s^{-1} sr^{-1} keV^{-1}$
- Quiet heliosphere: $0.035\ p\ cm^{-2} s^{-1} sr^{-1} keV^{-1}$

However, at present we ignore the fraction of time ATHENA will spend inside the different magnetotail zones (it will depend on the orbit radius among the other things). We assume that a fraction of time t>90% will be spent by the satellite outside the plasma sheet and adopt the following average flux at 80 keV: ~$10.5\ p\ cm^{-2} s^{-1} sr^{-1} keV^{-1}$.

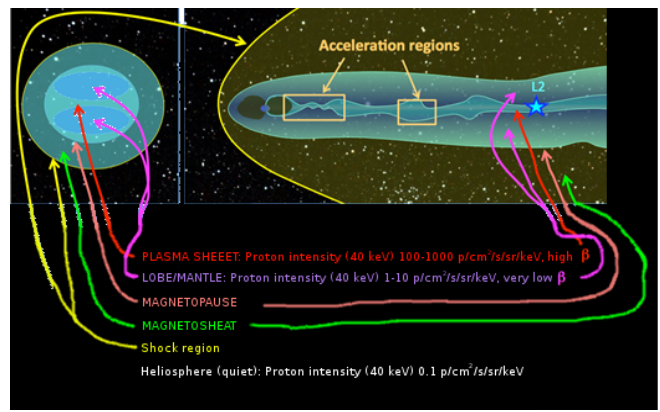

Figure 9. Schematic representation of the large scale magnetospheric structures/regions and their associated plasma regimes. The β plasma parameter is the ratio of the plasma thermal pressure over the magnetic pressure.

**Mirrors focalization efficiency**

The focalization efficiency of the ATHENA optics for soft protons has been calculated with two independent approaches:

- As level-0 approximation we can assume that the protons will have the same focalization efficiency of 1 keV photons:

$$\frac{n_{det}}{I_{inc}} \sim \frac{f^{xifu}_{cxb}(1\ keV)}{I^{ext}_{cxb}(1\ keV)} = \frac{\Omega(\theta) \cdot A_{opt}}{A_{det}} \sim 0.017\ sr$$

Where $f^{xifu}_{cxb}(1\ keV)$ is the flux of CXB photons impacting on the detector in $p\ cm^{-2}s^{-1}keV^{-1}$ and $I^{ext}_{cxb}(1\ keV)$ is the CXB intensity outside the optics, and $p\ cm^{-2}s^{-1}sr^{-1}keV^{-1}$, respectively.

- According to ray-tracing simulations on the ATHENA optics the focalization efficiency $\eta$ for protons is:

$$\eta = \frac{\Omega(\vartheta)}{4\pi} \frac{A_{opt}}{A_{det}} \frac{N_{det}}{N_{inc}} = 0.5 \times 10^{-3} \rightarrow \frac{n_{det}}{I_{inc}} = \eta \cdot 4\pi \sim 6.28 \times 10^{-3}\ sr$$

Where $\Omega(\theta)$ is the solid angle of the optics, $A_{opt}$ is the optics area in $cm^2$, $A_{det}$ is the detector area in $cm^2$, $n_{det} = \frac{N_{det}}{A_{det}}$ is the number of particles impacting on the detector per unit area in $p\ cm^{-2}$, and $I_{inc} = \frac{N_{inc}}{\Omega(\theta)A_{opt}}$ the intensity of the proton flux on the optics in $p\ cm^{-2}sr^{-1}$.

In the calculations we assumed $A_{det} = 2.4\ cm^2$ detector area, $A_{opt} = 20000\ cm^2$, $\Omega(\theta) = 2 \times 10^{-6}\ sr$.
The angular distribution of the focused proton beam will likely be Gaussian-shaped peaked towards the center of the FoV. The width of the distribution will depend on the proton energy.
The two estimates differ by a factor 3, confirming the validity of the estimate provided by the Monte Carlo code. In the following section we will use the ray-tracing outcome to estimate the flux at the focal plane.

**Magnetic diverter efficiency**

The magnetic diverter efficiency as function of the proton energy has been calculated by R. Willingale[7] with ray-tracing simulations, but only for few energies and relative to the Wide Field Imager FoV. At present we lack a real modelization for the X-IFU in the energy range of interest, so in the following we will use the upper limits on the fraction of transmitted protons for the WFI, scaling them by the ratio of the FoVs of the two instruments to obtain a conservative estimate for X-IFU (Figure 10).

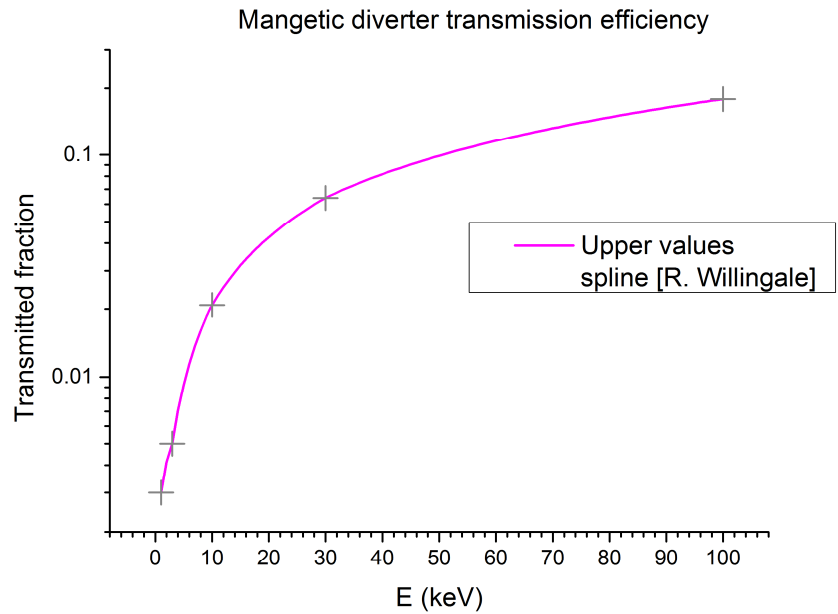

Figure 10. Soft protons fraction transmitted by the magnetic diverter as function of energy. The points are the upper values found for each energy with Monte Carlo simulations, and the line connecting them is a linear interpolation.

However, due to the lack of data regarding the efficiency for to the X-IFU, and to the poor sampling in the range 1-100 keV, these are to be considered just rough estimates.

**Proton energy losses**

In first approximation, we can conservatively assume that the protons do not lose energy in their interaction with the mirrors. The only energy loss will concern the radiation filters and the detector.

For XIFU the filters baseline currently consists of a total 0.28 um Kapton + 0.21 um Aluminum, divided in 5 filters of identical thickness. For such configuration, from Geant4 simulation we have the transmission function shown in Figure 11 – left, and the distribution of initial energies of protons that reach the focal plane with a residual energy in the range $0.2<E<10$ keV shown in Figure 11 – right, which can be taken as the transmission function for background-inducing protons.

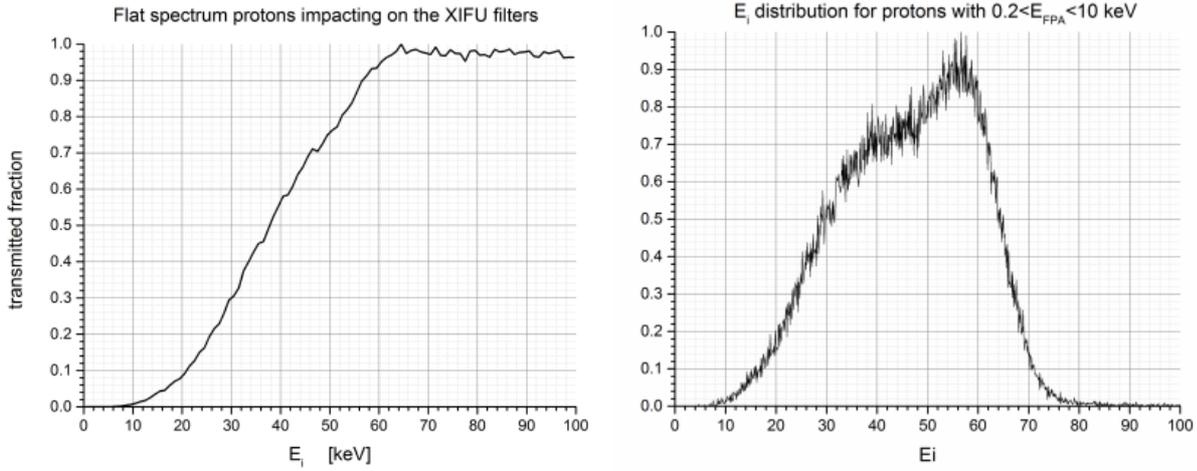

Figure 11. Transmission function for protons impacting on the X-IFU filters with a flat spectrum (left), and the initial energy distribution of protons that reach the focal plane with energy inside the X-IFU sensitivity band (right).

If we assume an impacting spectral shape $E^{-1.5}$ we can find the same distribution in a more realistic case (see Figure 12, left). Furthermore, from the corresponding cumulative distribution (Figure 12, right) it is easy to see that if we want to deflect 99% (99.9%) of protons that arrive at the focal plane with energies between 0.2 and 10 keV we must deflect protons with initial energies up to ~70 keV (80 keV).

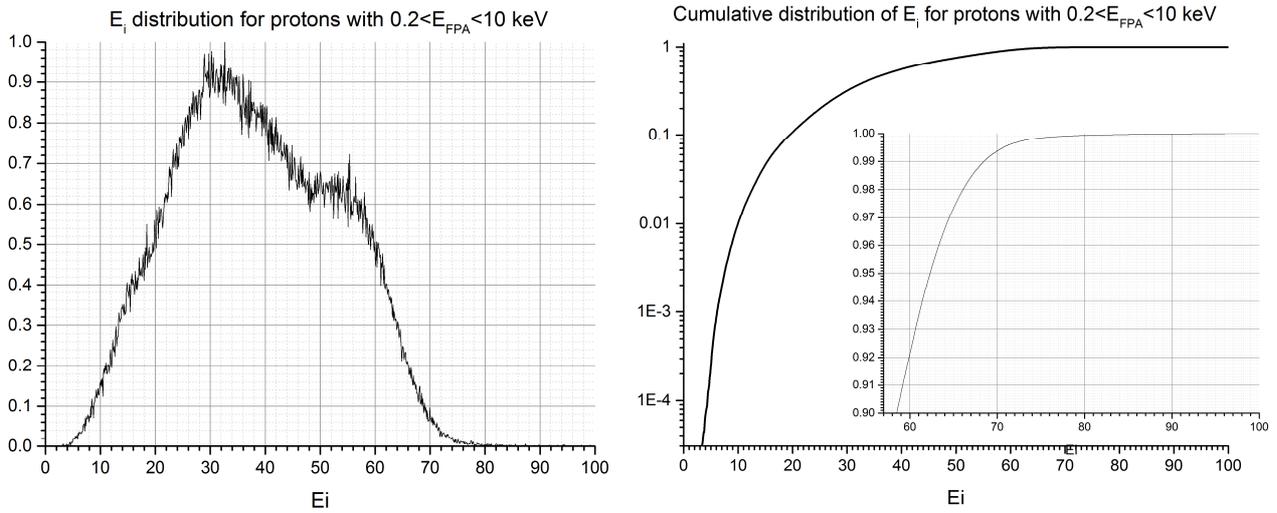

Figure 12. Distribution of initial energies of protons that reach the focal plane with energy inside the X-IFU sensitivity band, assuming an impacting spectral shape on the filters $E^{-1.5}$ (left), and the corresponding cumulative distribution (right).

The spectrum impacting on the filters however will be altered by the efficiency of the magnetic diverter. The effect of the transmission efficiency described in the previous section will be taken into account in the final computation. However, since the effect of the diverter is just to remove a fraction of the particles that would have impacted on the filters, the numbers derived in this section from Figure 11 – right and Figure 12 – right are conservative.

**Expected fluxes**

To evaluate the flux expected on the detector we have to fold the spectrum expected after the optics by the transmitted fractions, and by the transmission function of the filters (Figure 11 – right). The resulting spectra for particles that reach the FPA with energies inside the instruments sensitivity band are shown in Figure 13.

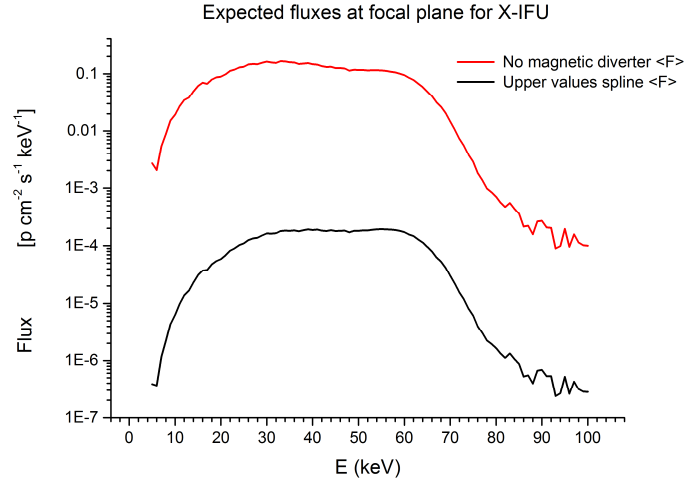

Figure 13. Distribution of initial energies of protons that reach the focal plane with energy inside the X-IFU sensitivity band, assuming an input spectrum $\propto E^{-1.5}$, the filters transmission functions derived in the previous section, with and without the magnetic diverter.

Integrating these spectra we find the following expected fluxes of particles: $F_{no\ magdiv} = 6.5\ p\ cm^{-2}\ s^{-1}$, $F = 8 \times 10^{-3}\ p\ cm^{-2}\ s^{-1}$ for the case without a magnetic diverter and using the upper values on its transmission, respectively. We want these fluxes to be $< 0.1 \times B_{GCR} = 5 \times 10^{-3}\ p\ cm^{-2}\ s^{-1}$, so the diverter should be able to reduce the flux of incoming soft protons by a factor ~1300. Our first conservative estimate of the efficiency of such diverter revealed that it brings the expected flux to the desired level.

The current estimate will benefit greatly from an improvement of the sampling of magnetic diverter efficiency in the 1-100 keV energy range with Monte Carlo simulations, from a series of simulations dedicated to the X-IFU instrument, and moreover from a more reliable estimate of the external fluxes of low energy particles expected in L2. The latter is expected as output of the AREMBES project.

### 1.3 Non-stationary conditions

During Solar Energetic Particle (SEP) events however the flux of high energy solar particles, usually negligible, can be enhanced by several orders of magnitude and reach the CR one, inducing additional background on the detector.

We analyzed the largest SEP events during solar cycle n° 23 (covering the time period 1997 – 2009), by using ACE/EPAM, SAMPEX and GOES/SEM data (available at http://www.srl.caltech.edu/sampex/_DataCenter/DATA/EventSpectra/). We selected sixteen SEP events extending to very high energies (> 100 MeV), which produce the so-called ground level enhancements (GLEs). The fluence spectrum for each event was fitted by using a broken power law functional form[8,9]. The differential flux (*dJ/dE*) as a function of the particle energy can be expressed as:

$$\frac{dJ}{dE} = \begin{cases} CE^{-\gamma_a} \exp\left(-E/E_0\right) & \text{for } E \leq (\gamma_b - \gamma_a)E_0 \\ CE^{-\gamma_b} \left\{ \left[(\gamma_b - \gamma_a)E_0\right]^{(\gamma_b - \gamma_a)} \exp(\gamma_a - \gamma_b) \right\} & \text{for } E > (\gamma_b - \gamma_a)E_0 \end{cases}$$

where $\gamma_a$ and $\gamma_b$ determine the spectral slope at energies lower and higher of the rollover energy $E_0$, respectively.

The left panel of Figure 14 shows the obtained fits for all the considered SEP events. It can be observed a variability of at least two orders of magnitude both at low and high energy. In addition, the spectral slopes are quite similar below the rollover energy, whereas they greatly differ at high energy. After verifying that the three sets of parameters are not mutually correlated between each other, we computed an average fluence spectrum, whose parameters ($\langle C \rangle$, $\langle \gamma_a \rangle$, $\langle \gamma_b \rangle$, $\langle E_0 \rangle$) are obtained as the average value of each parameter over the considered SEP events: $\langle C \rangle = 6.37*10^8$ p cm$^{-2}$ sr$^{-1}$ MeV$^{-1}$, $\langle \gamma_a \rangle = 1.24$, $\langle \gamma_b \rangle = 3.18$, $\langle E_0 \rangle = 29.45$ MeV). The average spectrum is displayed in the right panel of Figure 14 along with the one of the 4 April 2001 SEP event, which represents the worst case scenario at high energies.

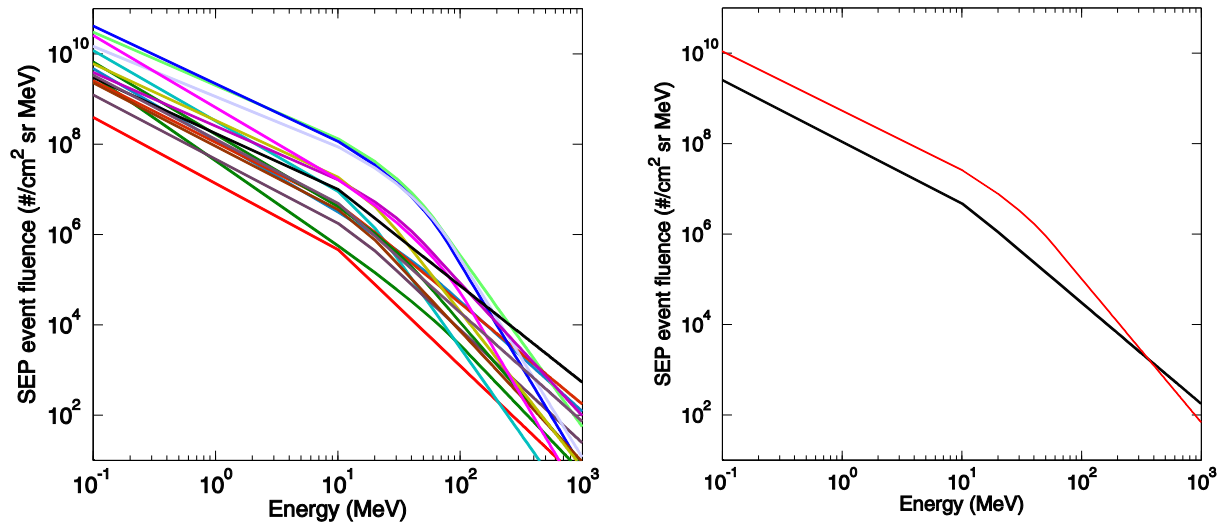

Figure 14. Left: Fits of the energy spectrum, by using the so-called broken power law function, for the SEP/GLE events during solar cycle n° 23. Right: Average spectrum (red line) and worst case scenario (black line, 4 April 2001 SEP/GLE event).

In order to estimate the occurrence of high energy SEP events, we analyzed the > 100 MeV proton flux data recorded aboard the GOES satellite series over two solar cycles from 1986 to 2009. We found that the fraction of time the > 100 MeV proton flux exceeds the GCR level (assumed to be 4.4 p cm$^{-2}$ s$^{-1}$) is about 1%, with a mean SEP event duration of ~1.06 days. This percentage slightly increases to 4% and 2% during the maximum phase of the solar cycle n° 22 (from 1988 to 1992) and n° 23 (from 1998 to 2002), respectively. Thus, these events are extremely unlikely, and the detectors will likely be shut down due to the even higher flux of low energy particles, so they pose no hindrance on the detector performances.

## 1.4 Conclusion

With the new X-IFU instrument and FPA mass models, and the Monte Carlo software version and settings, we obtained an updated estimate of the GCR induced particle background for the instrument. In the baseline configuration we expect 0.01 p cm$^{-2}$ s$^{-1}$ in the 2-12 keV energy band. Starting from this result we addressed the two main components of the unrejected background, secondary electrons and photons, and tested several solutions to reduce it.

Regarding the reduction of the photons component we tested several bi-layers and tri-layers for the passive shielding, in order to block fluorescences from the Niobium. The best result was obtained using a bilayer made of 250 μm of Kapton and 1.3 mm of $Si_3N_4$, roughly halving the photon component and reducing the total background by ~25%.

Regarding the secondary electrons component, we tested the performances of a composite shield, substituting the lowest section of the Kapton shield with Tungsten (1 cm high, 1 mm thick). This resulted in a background reduction of ~20%. The other solution tested, the insertion of a filter right above the detector, resulted in too high thicknesses required for such filter to reduce the background by a significant amount. The reliability of this last result is yet to be confirmed by the Geant4 processes validation activity foreseen in AREMBES.

We also performed a first estimate of the background induced by low energy particles impacting on the Athena optics. We found that without any magnetic diverter we expect a focused particles flux $F_{no\ magdiv} = 6.5\ p\ cm^{-2}\ s^{-1}$, several OoM higher than the requirement. Using the upper values on the magnetic diverter transmitted fraction we obtained an expected flux $F = 8 \times 10^{-3}\ p\ cm^{-2}\ s^{-1}$. The requirement for the soft protons component of the background is for it to be $< 0.1 \times B_{GCR} = 5 \times 10^{-3}\ p\ cm^{-2}\ s^{-1}$, so the diverter should be able to reduce the flux of incoming soft protons by a factor ~1300. Our first conservative estimate of the efficiency of such diverter revealed that it brings the expected flux to the desired level. The magnetic diverter transmission values we used are upper limits scaled from the WFI FoV, and assume an isotropic distribution of the protons at the focal plane, and thus this first estimate is to be considered conservative for X-IFU.

Finally, we have analyzed data from particle monitors in different locations in the solar system, with the aim to quantify the impact of solar events on the background, and found that the fraction of time the flux exceeds the GCR level is about 1%, with a mean SEP event duration of ~1.06 days. This percentage did not exceed 4% even during the maximum phase of the solar cycle, thus we concluded that these events can be safely ignored.

## ACKNOWLEDGEMENT

This work has been supported by ASI (Italian Space Agency) through the Contract n. 2015-046-R.0. The authors acknowledge also CNES, IRAP and SRON for the provided material.

**ACKNOWLEDGE ALSO CNES, IRAP AND SRON FOR THE PROVIDED MATERIAL.**